\documentclass[conference]{IEEEtran}
\IEEEoverridecommandlockouts
\usepackage{cite}
\usepackage{amsmath,amssymb,amsfonts}
\usepackage{algorithmic}
\usepackage{graphicx}
\usepackage{textcomp}
\usepackage{xcolor}
\usepackage{soul}
\def\BibTeX{{\rm B\kern-.05em{\sc i\kern-.025em b}\kern-.08em
    T\kern-.1667em\lower.7ex\hbox{E}\kern-.125emX}}
\begin{document}

\title{Automatic Lung Cancer Prediction from Chest X-ray Images Using Deep Learning Approach}
\author{\IEEEauthorblockN{Worawate Ausawalaithong}
\IEEEauthorblockA{\textit{Kamnoetvidya Science Academy} \\
Rayong, Thailand \\
a.worawate@gmail.com}\\
\IEEEauthorblockN{Sanparith Marukatat}
\IEEEauthorblockA{\textit{National Electronics and Computer Technology Center}\\
Patumthani, Thailand \\
sanparith.marukatat@nectec.or.th}
\and
\IEEEauthorblockN{Arjaree Thirach}
\IEEEauthorblockA{\textit{Kamnoetvidya Science Academy} \\
Rayong, Thailand \\
arjaree.t@kvis.ac.th}\\
\IEEEauthorblockN{Theerawit Wilaiprasitporn}
\IEEEauthorblockA{\textit{School of Information Science and Technology} \\
\textit{Vidyasirimedhi Institute of Science and Technology}\\
Rayong, Thailand \\
theerawit.w@vistec.ac.th}
}

\maketitle

\begin{abstract}
Since, cancer is curable when diagnosed at an early stage, lung cancer screening plays an important role in preventive care. 
Although both low dose computed tomography (LDCT) and computed tomography (CT) scans provide more medical information than normal chest x-rays, there is very limited access to these technologies in rural areas. 
Recently, there is a trend in using computer-aided diagnosis (CADx) to assist in screening and diagnosing of cancer from biomedical images. 
In this study, the 121-layer convolutional neural network also known as DenseNet-121 by G. Huang et. al., along with the transfer learning scheme was explored as a means to classify lung cancer using chest X-ray images. 
The model was trained on a lung nodules dataset before training on the lung cancer dataset to alleviate the problem of a small dataset. 
The proposed model yields 74.43$\pm$6.01\% of mean accuracy, 74.96$\pm$9.85\% of mean specificity, and 74.68$\pm$15.33\% of mean sensitivity. 
The proposed model also provides a heatmap for identifying the location of the lung nodule. 
These findings are promising for further development of chest x-ray-based lung cancer diagnosis using the deep learning approach. 
Moreover, these findings solve the problem of small dataset.
\end{abstract}

\section{Introduction}
{As reported by WHO, cancer has caused a very large number of deaths, approximately 8.8 million deaths in 2015 \cite{WHOstat}.
Almost 20\% or 1.69 million deaths are due to {\em lung cancer}\cite{WHOstat}. 
Cancer screening plays an important role in preventive care because cancer is most treatable when caught in the early stages. 
The study shows that the appearances of the malignant lung nodules more commonly demonstrate a spiculated contour, lobulation, and inhomogeneous attenuation \cite{PulmonaryNodule}. 

\indent
Presently, low dose computed tomography (LDCT) plays an important role in screening lung cancer \cite{RuralAreas}. 
LDCT screening has reduced lung cancer deaths and is recommended for high-risk demographics \cite{RuralAreas}. 
Results from LDCT screening may be further evaluated with standard dose computed tomography (CT)  \cite{CTscan}. 
However, there are many barriers to implementing LDCT screening, such as providers' anxiety on the access to LDCT equipment and potential financial burdens in rural populations \cite{RuralAreas}. 
Moreover, rural populations have limited access to both primary care physicians and specialists \cite{RuralAreas}. 
On the other hand, chest x-rays are readily available in rural areas.
Nonetheless, chest x-rays produce lower quality images compared to LDCT or CT scans and, therefore, a lower quality diagnosis is generally expected.
This study explores the use of chest x-rays with a computer-aided diagnosis (CADx) system to improve lung cancer diagnostic performance. 

\indent
The convolutional neural network (CNN) is proven to be very effective in image recognition and classification tasks. 
The development of CNNs starts from, LeNet \cite{LeNet}, AlexNet \cite{AlexNet}, ZFNet \cite{ZFNet}, VGG \cite{VGG}, Inception \cite{Inceptionv1} \cite{Inceptionv23}, ResNet \cite{ResNet}, Inception-ResNet \cite{InceptionResNet}, Xception \cite{Xception}, DenseNet \cite{DenseNet}, and NASNet \cite{NASNet}. 
There are many studies on the use of CNNs to detect abnormalities in chest x-rays using the deep convolutional neural network. 
For instance, M. T. Islam et al., \cite{AbnormDetection} used several CNNs to detect abnormalities in chest x-rays. 
There is also a study, by X. Wang et al., on using a couple of CNNs to detect thoracic pathologies from chest x-ray images.
This study also provides a large dataset that was used in our study \cite{ChestX-ray14}. 
Among the research, there are some studies about applying Densely Connected Convolutional Networks (DenseNet) \cite{DenseNet} to detect thoracic pathologies such as ChexNet \cite{ChexNet} and the Attention Guided Convolutional Neural Network (AG-CNN) \cite{AGCNN}. 
Both studies trained the neural network on a very large chest x-ray image dataset. 

\indent
Lung cancer prediction with CNN faces the small sample size problem.
Indeed, CNN contains a large number of parameters to be adjusted on large image dataset.
In practice, researchers often pre-trained CNNs on {\em ImageNet}, a standard image dataset containing more than one million images.
Then, the trained CNNs are adjusted on a specific target image dataset.
Unfortunately, the available lung cancer image dataset is too small for this transfer learning to be effective, even with a data augmentation trick.
To alleviate this problem, we proposed the idea of applying transfer learning several times to gradually improve the performance of the model. 
In this work, the transfer learning is applied twice.
The first time is to transfer the model from general image domain into chest x-ray domain.
The second time is to transfer the model to lung cancer.
This multi-transfer learning can solve the small sample size problem and achieves a better result on the task compared to the traditional transfer learning technique. 

Furthermore, we also show that it is possible to modify the model to compute heatmap showing the predicted position of lung cancer in the chest x-ray image.}

\section{Methodology}
{In this section, the methodology of our study is discussed. 
First a description of the datasets used in the study is given, followed by a description the data preparation process. 
Consequently, an explanation is given of the model architecture and the loss function, and finally the visualizing process.}
\subsection{Datasets}
\subsubsection{JSRT Dataset \cite{JSRT}}
{This public dataset from JSRT (Japanese Society of Radiological Technology) consists of 247 frontal chest x-ray images, of which 154 images have lung nodules (100 malignant cases, 54 benign cases) and 93 are images without lung nodules. 
All images have a size of $2048\times 2048$ pixels. 
Since, this dataset is very small, a 10-fold cross-validation was performed to increase reliability on the performance. 
Malignant cases were used as positive cases, while both benign and non-nodule cases were used as negative cases. 
Data were randomly split into the training, validation, and test set, with the ratio shown in Table \ref{table: 1}.}
\subsubsection{ChestX-ray14 Dataset \cite{ChestX-ray14}}
{This public dataset released by Wang et al. \cite{ChestX-ray14} contains 112,120 frontal chest x-ray images, currently the largest public chest x-ray dataset. Each image is individually labeled with up to 14 different thoracic pathology labels. All images have a size of $1024\times 1024$ pixels. However, this dataset does not contain any lung cancer images. The fourteen thoracic pathology labels consist of Atelectasis, Consolidation, Infiltration, Pneumothorax, Edema, Emphysema, Fibrosis, Effusion, Pneumonia, Pleural Thickening, Cardiomegaly, Nodule, Mass, and Hernia. It was decided to use this dataset to compensate the lung cancer data that has only 100 cases, by first training for the recognition of nodules, using nodule cases as positive and all remaining cases (non-nodule) as negative. Data were randomly split into the training, validation, and test set, without having any overlap of patients between the data sets, as shown in Table \ref{table: 1}.}
\begin{table}
\caption{Training-Validation-Test split for each dataset}
\label{table: 1}
\centering
\begin{tabular}{cccc}
\hline
& Training & Validation & Test\\ \hline
JSRT Dataset & 80 & 10 & 10\\
\hline
ChestX-ray14 Dataset & & &\\
Positive cases & 4992 & 1024 & 266\\
Negative cases & 103907 & 1024 & 266\\
Total & 108899 & 2048 & 532\\
\hline\hline
\end{tabular}
\end{table}

\subsection{Data Preparation}
\begin{figure}
\includegraphics[width=1.0\linewidth]{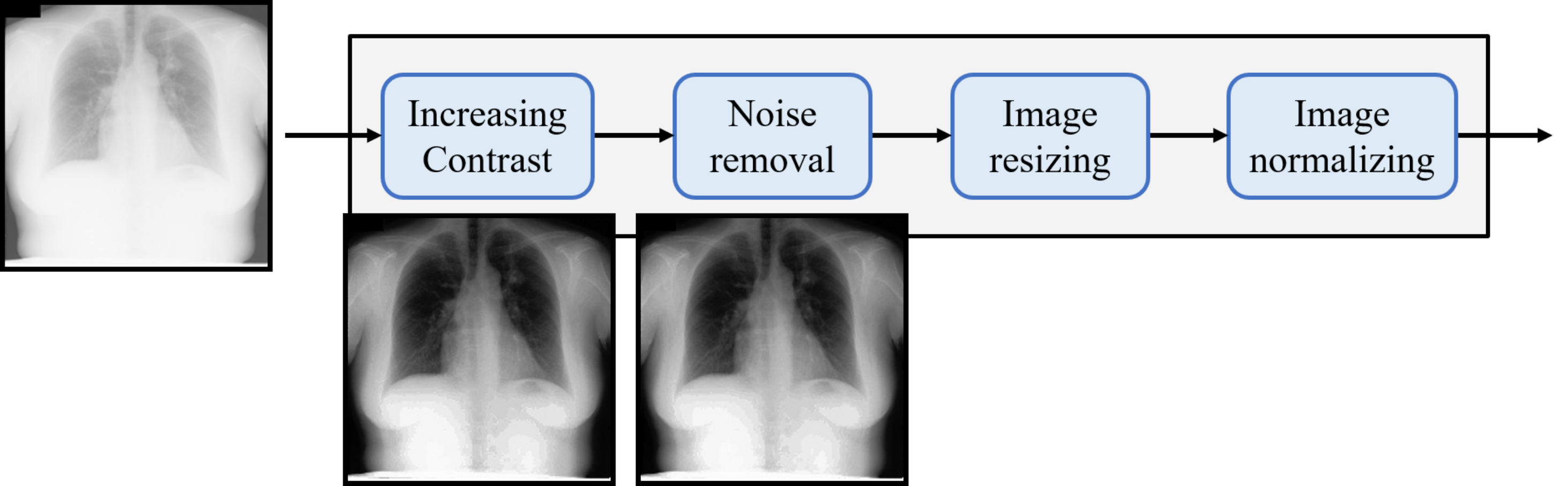}
\centering
\caption{Illustration of Data Preparation process on example image from JSRT dataset.}
\label{Fig1}
\end{figure}

{Data preparation was applied to all the images. It consists of four steps as follows: 
\begin{itemize}
\item Step 1: increasing the contrast of all images using Histogram Equalization. This allows normalizing the intensity of images from different datasets.
\item Step 2: removing the noise from all images using Median Filtering with a window size of $3\times 3$.
\item Step 3: resizing images to $224\times 224$ pixels to match the input of the model used in this study.
\item Step 4: normalizing image color based on mean and standard deviation of the ImageNet training set \cite{ImageNet}.
\end{itemize}
An example of the image after processing in each step is shown in Fig. \ref{Fig1}.}

\subsection{CNN architecture and transfer learning}
{Among all the convolutional neural networks architectures like Inception \cite{Inceptionv1} \cite{Inceptionv23}, ResNet \cite{ResNet}, DenseNet \cite{DenseNet}, the model we used in this research is the 121-layer Densely Connected Convolutional Networks (DenseNet-121).
This convolutional neural network has the dense blocks architecture that improves the flow of the data along the network, and also solves the problem of vanishing gradient found in very deep neural networks. 
This CNN also performs very well on many public datasets including ImageNet \cite{DenseNet}.
The original model processes an input image of size $224\times 224$ pixels and outputs posterior probabilities for 1000 categories of object.
In this work, the last fully connected layer of this CNN was replaced by a single sigmoid node in order to output the probability of having the specified pathology. 

{Due to the small dataset problem, we considered transfer learning.
However, as the lung cancer dataset is truly small, transfer learning was applied more than usual. 
Specifically, in this study, it was applied twice. 
The first transfer learning allowed the classification of chest x-ray images as "with nodule" or "without nodule". 
The second allowed the classification of chest x-ray images as "with malignant nodule" and "without malignant nodule" (including both chest x-rays with benign nodules and those without nodules).
Through the training process, the model became more specific to the lung cancer task. 
The overall training process and the model used in each step is described below and in Fig. \ref{Fig3}, overleaf:
\begin{figure}
\includegraphics[width=0.9\linewidth]{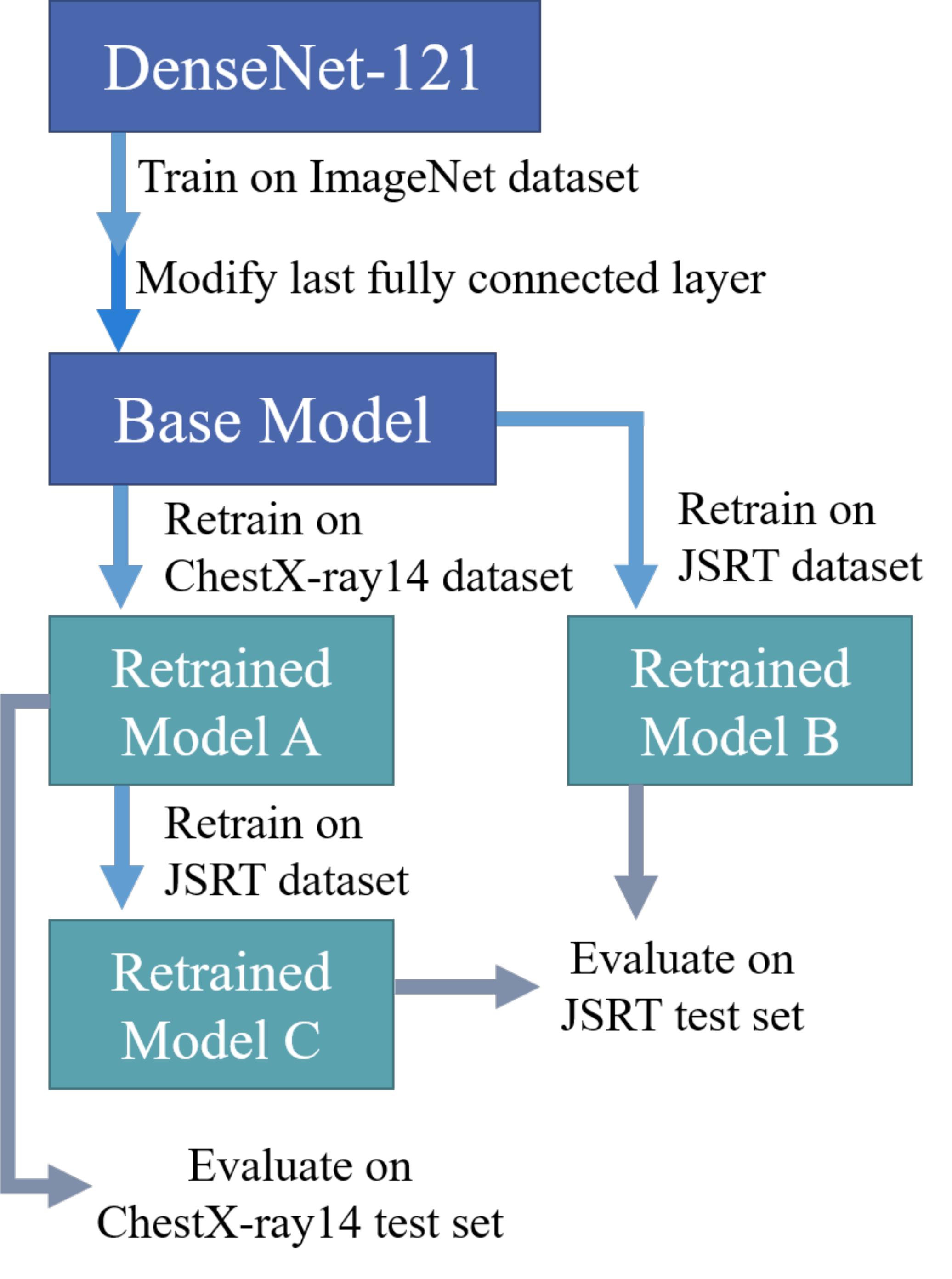}
\centering
\caption{The training process and the model in use including Base Model, Retrained Model A, B, or C, along with the data used to train each model.}
\label{Fig3}
\end{figure}
\begin{itemize}
\item Base Model: DenseNet-121 with the initial weight from pretrained model on ImageNet training set that has the last fully connected layer modified to 1 class of sigmoid activation function.
\item Retrained Model A: Retrained Base Model with the least validation loss from the ChestX-ray14 validation set. Base Model was retrained using the ChestX-ray14 training set, as shown in Table \ref{table: 1}, where positive cases were chest x-ray images with nodules and negative cases are the chest x-ray images without nodules. The training images were randomly flipped horizontally.
\item Retrained Model B: Retrained Base Model with the least validation loss from the JSRT validation set. The Base Model was retrained using the JSRT training set, as shown in Table \ref{table: 1}, along with a 10-fold cross-validation. The training images were randomly rotated by the angle within 30 degrees and randomly flipped horizontally. Positive cases were chest x-ray images with malignant nodules, while negative cases were chest x-ray images without malignant nodules.
\item Retrained Model C: Retrained Model A with the least validation loss from the JSRT validation set. Due to the fact that the positive cases of the JSRT dataset are all nodules, the Retrained Model A, (which already had the ability to classify "nodule" from "non-nodule", a task similar to the lung cancer classification task) was retrained to identify malignant vs. non-malignant nodules. The JSRT training set was used to retrain the model. All the JSRT training images used to retrain the model were randomly rotated within 30 degrees and randomly flipped horizontally.
\end{itemize}

\indent The basic idea for the training of Model C is as follows:
First, DenseNet-121 was trained on the ImageNet dataset to gain basic knowledge about images in general.
Next, the model was retrained on the ChestX-ray14 dataset to adjust this knowledge to chest x-ray images with additional information about lung nodules.
As the size of the ChestX-ray14 dataset is much larger than the JSRT dataset, transferring knowledge to the chest x-ray domain first allows better adaptation of the CNN that has been trained on a general images dataset.
Model A can be considered as a chest x-ray specialist.
Additionally, as lung nodules are very similar to lung cancer, Model A is ready to be adjusted using the JSRT dataset.
Experimentally, we have found that Model C shows the best performance compared to single-step transfer learning (i.e. Model B).
This underlines the effectiveness of the proposed strategy.

\subsection{Loss and optimizer}
\indent {The two transfer learnings involve imbalanced binary classification datasets. 
To counteract this problem, the following weighted binary classification loss was used:
\begin{eqnarray}
L(X,y)&=&-\omega_+ \cdot y\log p (Y=1|X)\notag\\&&-\omega_- \cdot (1-y) \log p (Y=0|X),
\end{eqnarray}
where $X$ is the image and $y$ is the real label of the image which is labeled 0 if being a negative case, or 1 if being a positive case, $p(Y=i|X)$ is the predicted probability of having label $i$. $\omega$ is the weight applied on the loss to make the training process more efficient, $\omega_+$ is the number of negative cases per all cases and $\omega_-$ is the number of positive cases per all cases.}

\indent {Adam (adaptive moment estimation) optimizer \cite{Adam} was used in this work with the standard parameter setting, i.e. $\beta_{1} = 0.9$ and $\beta_{2} = 0.999$. 
The initial learning rate was 0.001 and decreased by a factor of 10 when validation loss plateaued. 
The batch size was 32.}

\subsection{Class Activation Mappings (CAMs)}
Class Activation Mappings (CAMs) \cite{CAMs} can be derived from Model C to show the most salient location on the image that the model used to identify the output class.
This is done using the following equation: 
\begin{eqnarray}
M_c=\sum_{k}w_{c,k}f_k.
\end{eqnarray}
From equation (2), $M_c$ is the map that shows the most salient features on the image that the model used to classify classes. $w_{c,k}$ is the weight in the last fully connected layer of the feature map $k$ leading to class $c$ and $f_k$ is the $k$th feature map.

\begin{table*}
\caption{The name of retrained model, dataset used, purpose, accuracy, specificity, and sensitivity of Retrained Model A, B, and C on tasks}
\label{table: 2}
\centering
\begin{tabular}{cccccc}
\hline
Retrained Model & Dataset used & Purpose & Accuracy & Specificity & Sensitivity\\ \hline
A & ChestX-ray14 & to recognize lung nodules & 84.02\% & 85.34\% & 82.71\%\\
B & JSRT & to recognize lung cancer & $65.51\pm 7.67$\% & $80.95\pm 20.59$\% & $45.67\pm 21.36$\%\\
C & ChestX-ray14 and JSRT & to recognize lung cancer & $74.43\pm 6.01$\% & $74.96\pm 9.85$\% & $74.68\pm 15.33$\%\\
\hline\hline
\end{tabular}
\end{table*}

\section{Results and Discussion}

\begin{figure}
\includegraphics[width=0.9\linewidth]{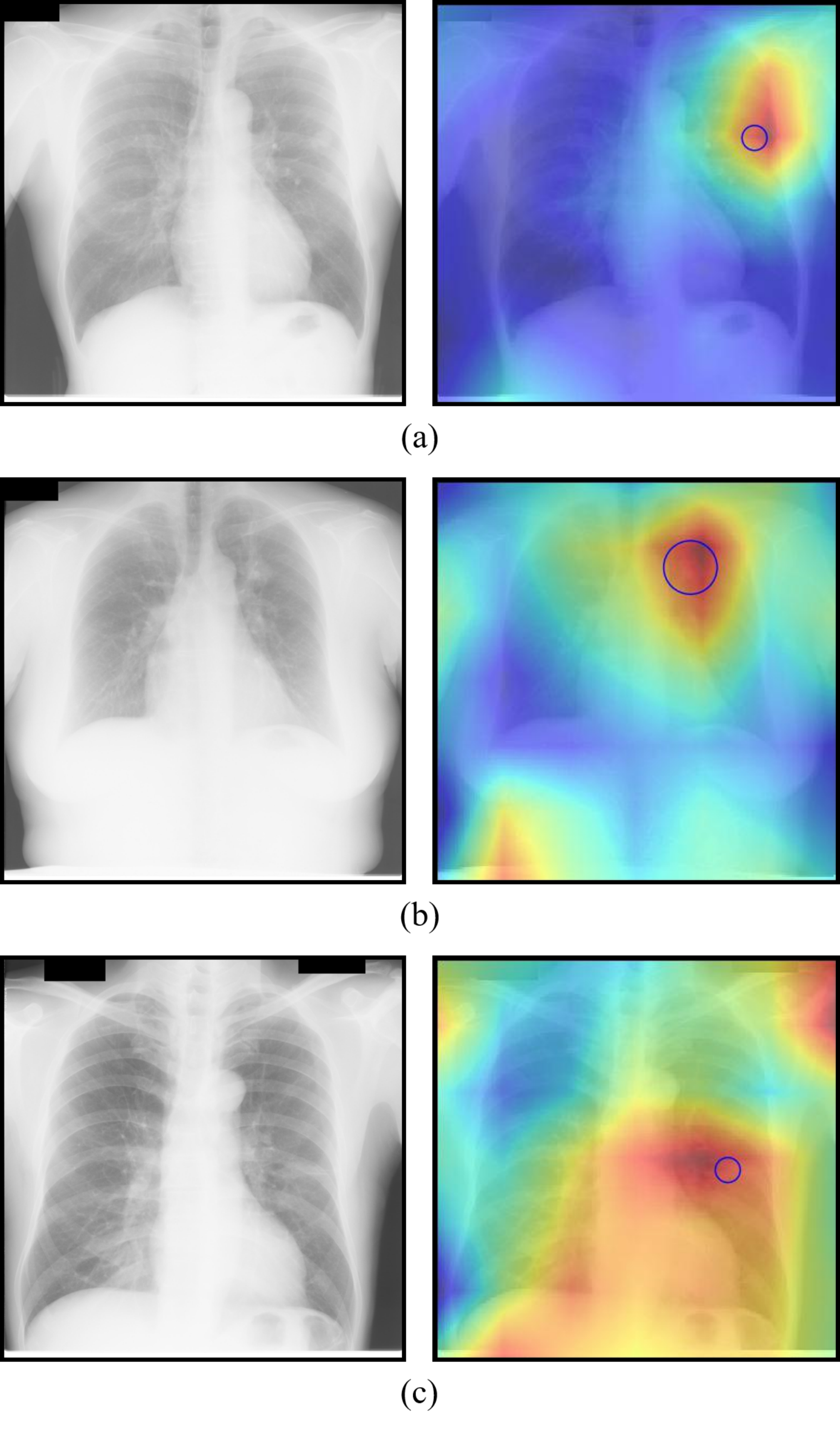}
\centering
\caption{Images of malignant cases from JSRT dataset, before and after illustrate Class Activation Maps on using Retrained Model C.}
\label{Fig2}
\end{figure}

{The performance of the model in this study was evaluated using accuracy, specificity, and sensitivity. The accuracy shows the degree to which the model correctly identified both positive and negative cases. The specificity shows the degree to which the model correctly identified negative cases and sensitivity shows the degree to which the model correctly identified positive cases. By having high accuracy, specificity and sensitivity, it can be implied that the model has low error. The performance of Model A on classifying lung nodules was evaluated for accuracy, specificity and sensitivity using a test set of the ChestX-ray14 Dataset as shown in Table \ref{table: 2}. The threshold of Model A is 0.55 as it gives highest sum between specificity and sensitivity. The performance of Model B and Model C on classifying lung cancer were likewise evaluated using the average and standard deviation of accuracy, specificity and sensitivity using a test set of the JSRT dataset in each fold of a 10-fold cross-validation as shown in Table \ref{table: 2}. The threshold of both models is 0.5.}

{As shown in Table \ref{table: 2}, Model A, which was trained to recognize lung nodule, after being pretrained on the ImageNet dataset, performed efficiently on the test set of the ChestX-ray14 dataset. Next, Model B resulted in higher specificity but poorer in both accuracy and sensitivity than Model C. In addition, Model C also has lower standard deviation in all evaluation metrics. The result shows that by retraining the model several times for specific tasks gives better results in almost all metrics.}

{In Fig.\ref{Fig2}, left hand images are the original images while the right hand images are processed images, using CAMs to show the salient location on the image, with the blue circle corresponding to the actual location of lung cancer on the image, identified by information received with the JSRT images. Model C can show accurate CAMs in most of the correctly predicted images such as (a) and (b), but some images show too large an area over the actual location or show an inaccurate area. The reason is that the model still overfits the training set slightly due to the size of the dataset. On the other hand, CAMs illustrated by Model B do not show the accurate position of lung cancer and often show that the model used too large an area on the image to classify lung cancer.}

\section{Conclusion}
{To conclude, the performance of the densely connected convolutional network on detecting lung cancer from chest x-ray images was explored. Since the dataset was too small to train the convolutional neural network, was proposed a strategy to train the deep convolutional neural network using a very small dataset. The strategy was to train the model several times, having the model learn about the final task step-by-step, which in this case, starts from the use of general images in the ImageNet dataset, followed by identifying nodules from chest x-rays in ChestX-ray14 dataset, and finally identifying lung cancer from nodules in the JSRT dataset. The proposed training strategy performed better than the normal transfer learning method, which resulted in higher mean accuracy and mean sensitivity but poorer mean specificity, as well as a lower standard deviation. The region used to classify images, was illustrated using CAMs in Model C, showing quite accurate locations of lung cancer, although some do not due to the overfit problem. However, CAMs from Retrained Model B are much poorer. Therefore, it can be concluded that the proposed method can solve the problem of a small dataset.}
\newline \indent
{For future work, the data augmentation can be additionally performed by randomly adding Gaussian Noise, and randomly cropping the image. The idea of Attention-Guided Convolutional Neural Network (AG-CNN) \cite{AGCNN} can also be used to crop the nodule part from the full image and use its real size to identify the malignancy, since the appearance of malignant nodules is different from benign nodules. Additionally, more features such as family history, and smoking rate can be applied in training along with images for more accuracy. The ensemble of several convolutional neural networks can further improve the results.}
\bibliographystyle{IEEEtran}
\bibliography{ref} 
\end{document}